\documentclass{iau} 
\usepackage{graphicx}
\usepackage{url}

\title[Emission-line Diagnostics of HII Regions] 
{Emission-line Diagnostics of Nearby HII Regions Including Supernova Hosts}

\author[Lin Xiao et al.]   
{Lin Xiao$^1$,
 J.J. Eldridge$^2$, Elizabeth Stanway$^3$, and L. Galbany$^4$}
\affiliation{$^1$,$^{2}$Department of Physics, University of Auckland, NZ\\ email: {\tt lin.xiao@auckland.ac.nz, j.eldridge@auckland.ac.nz} \\[\affilskip]
$^3$Department of Physics, University of Warwick, Gibbet Hill Road, Coventry, CV4 7AL, UK \\email: {\tt e.r.stanway@warwick.ac.uk}\\
[\affilskip]
$^4$PITT PACC, Department of Physics and Astronomy, University of Pittsburgh, Pittsburgh, PA 15260, USA \\email: {\tt lgalbany@das.uchile.cl}
}

\pubyear{2015}
\volume{xxx}  
\jname{Title of your IAU Symposium}
\editors{A.C. Editor, B.D. Editor \& C.E. Editor, eds.}
\begin{document}

\maketitle

\begin{abstract}
We present a new model of the optical nebular emission from HII regions by combining the results of the Binary Population and Spectral Synthesis (\textsc{bpass}) code with the photoionization code \textsc{cloudy} (\cite[Ferland et al. 1998]{1998PASP..110..761F}). We explore a variety of emission-line diagnostics of these star-forming HII regions and examine the effects of metallicity and interacting binary evolution on the nebula emission-line production. We compare the line emission properties of HII regions with model stellar populations, and provide new constraints on their stellar populations and supernova progenitors. We find that models including massive binary stars can successfully match all the observational constraints and provide reasonable age and mass estimation of the HII regions and supernova progenitors. 
\end{abstract}

The spectra of star-forming HII regions carry a wealth of information about properties of gas cloud and stellar populations. A number of emission line features in a spectrum allow us to derive element abundances and set valuable constraints on the age and mass of stellar population (e.g., \cite[Byler et al. 2016]{2016arXiv161108305B}; \cite[Levesque et al. 2010]{2010AJ....139..712L}) and supernova (SN) progenitors (\cite[Galbany et al. 2016]{2016A&A...591A..48G}). These studies all assume single star models, the inclusion of interacting binaries in stellar populations is a major step forward in modelling populations accurately (e.g. \cite[Stanway et al. 2016]{2016MNRAS.456..485S}). Here we discuss the importance of including interacting binaries in modelling HII regions.
\firstsection 
                       
\section{HII Regions Sample}
We consider two surveys of HII regions - one isolating such regions in nearby galaxies, and a second identifying the emission line properties at the location of known core-collapse supernovae (CCSNe). We study these separately to place constraints on massive star populations and CCSN progenitors.

First we use the van Zee sample of nearby individual HII regions. There are a total of 254 HII regions with 188 from 13 spiral galaxies (\cite[van Zee et al., 1998]{1998AJ....116.2805V}) and 66 from 21 dwarf galaxies (\cite[van Zee \& Haynes 2006]{2006ApJ...636..214V}). The one-dimensional spectra of these HII regions were corrected for atmospheric extinction, flux-calibrated and optical emission line fluxes were measured relative to H$ {\rm \beta} $ intensity. In the work of \cite[van Zee et al. (1998)]{1998AJ....116.2805V}, abundances for several elements (oxygen, nitrogen, neon, sulfur, and argon) were determined and radial abundance gradients were derived for 11 spiral galaxies. 

The second sample comes from the PMAS/PPAK Integral-field Supernova hosts COmpilation (PISCO, Galbany et al. in prep). It is composed of observations of SN host galaxies selected from the CALIFA survey and other programs using the PMAS/PPAK Integral Field Unit with large FoV and a spatial resolution of 1 arcsec. \cite[Galbany et al.(2014)]{2014A&A...572A..38G} have compiled 128 SNe and, for the first time, spectroscopically probed the association of different SN types to different formation environments. Moreover, the differences in the local environmental oxygen abundance for different SNe have been also investigated and they find the location of SNe Ic and II show higher metallicity than those of SNe Ib, IIb, and Ic-BL (\cite[Galbany et al. 2016]{2016A&A...591A..48G}).

\section{Nebular Emission Model}
To reproduce the observed nebular emission lines we combine \textsc{bpass} v2.0 stellar population models described in detail in \cite[Stanway et al., (2016)]{2016MNRAS.456..485S} and reference therein with the photoionization code \textsc{cloudy} (\cite[Ferland et al., 1998]{1998PASP..110..761F}). In brief, the BPASS models are a set of publicly available stellar population synthesis models which are constructed by combining stellar evolution models with the latest stellar synthetic spectral models. Our aim is to investigate how binary evolution, which dominates in star-forming regions, can differ from single evolution in terms of nebular emission production. 

Binary interactions lead to a variety of evolutionary pathways, such as preventing stars from becoming red giants by removing their hydrogen envelopes. These stars then become helium-rich dwarfs and evolve at much higher surface temperatures. Also a companion star can accrete material or two stars can merge, creating a star that is more massive than it was initially. This can upset the simple relationship between the most massive or luminous star and the age of the population of which it is a member.

Interacting binaries thus can lead to very different input ionizing spectra for \textsc{cloudy} models. In our work, we assume the gas nebula without any dust to be spherical, ionization-bound and to have a constant, non-evolving hydrogen density $ {\rm n(H)} $ spanning logarithmically from 0 to 3 with 0.5 dex intervals. Moreover, we consider 13 values of Z between 0.05 per cent of and 2 times solar (Z=0.020), corresponding to metallicities at which stellar population models are available. We compute models for 21 values of U logarithmically spaced in the range from $ -3.5 $ to $ -1.5 $, in steps of 0.1 dex at 21 stellar population ages between 1 and 100 Myrs in 0.1 dex bin. In addition, we specify the 9 essential elements' abundance of the nebula in terms of metallicity, Z, with $ {\rm Hydrogen = 0.75 - 2.5 \times Z }$; $ {\rm Helium = 0.25 +1.5 \times Z }$; $ {\rm Carbon = 0.173 \times Z }$; $ {\rm Oxygen = 0.482 \times Z }$; $ {\rm Nitrogen = 0.053 \times Z }$; $ {\rm Neon = 0.099 \times Z }$; $ {\rm Magnesium = 0.038 \times Z }$; $ {\rm Silicon = 0.083 \times Z }$; $ {\rm Iron = 0.072 \times Z }$.  
 
As a consequence, we obtain thousands of photoionization models and find the spectral difference between single and binary model varies with stellar age. At stellar ages of less than 5 Myrs, both single-star and binary-star models produce almost same UV radiation and emit the same  strong emission lines. However, at ages up to 10 Myrs, there is more UV radiation produced by binary-star models than single-star models. Therefore the population still emits emission lines that are lost from single-star models. This is due to production of hot helium stars or WR stars in the binary models which are not made in the single-star population due to the limited strength of stellar wind mass loss. More detailed discussion can be found in Stanway et al. (2016) which investigates the relationship between stellar population and ionizing flux, and is summarised elsewhere in this proceedings (Stanway 2017). Therefore, with these binary-star population to heat the nebulae, the HII regions can survive longer than 10 Myrs, assuming the interstellar medium remains sufficiently dense.

\section{Comparison to Observations}
\begin{figure}[h]
\begin{center}
\hspace*{-1.5 cm} \includegraphics[width=6.5in]{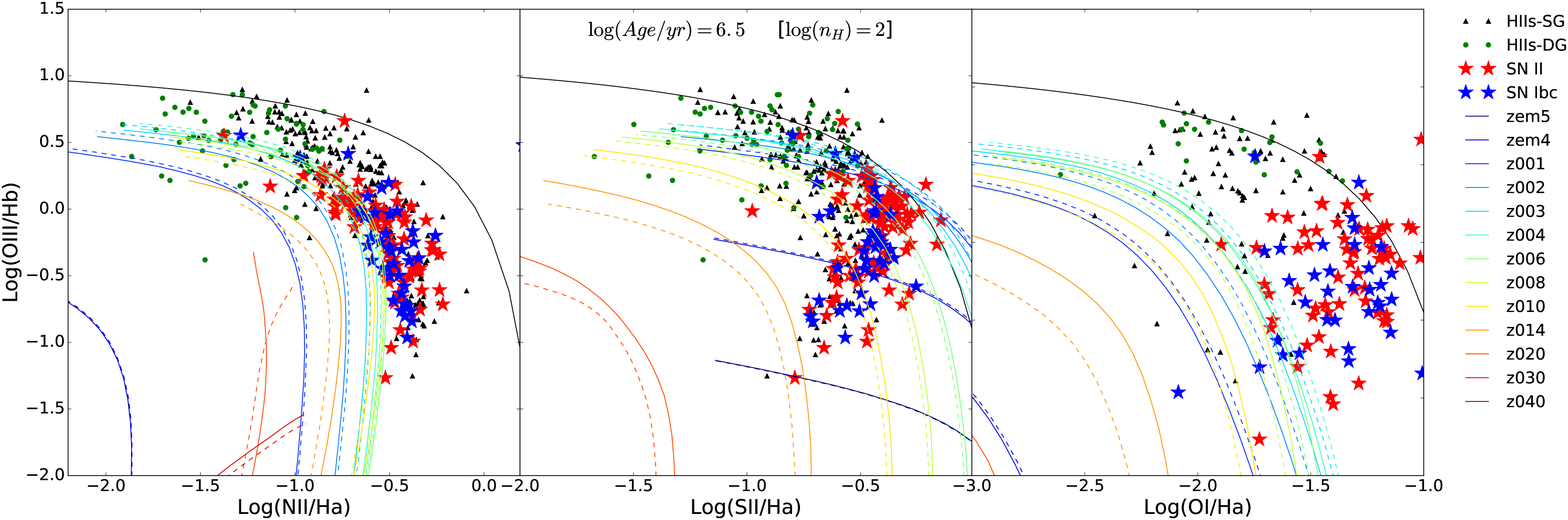} 
\hspace*{-1.5 cm}\includegraphics[width=6.5in]{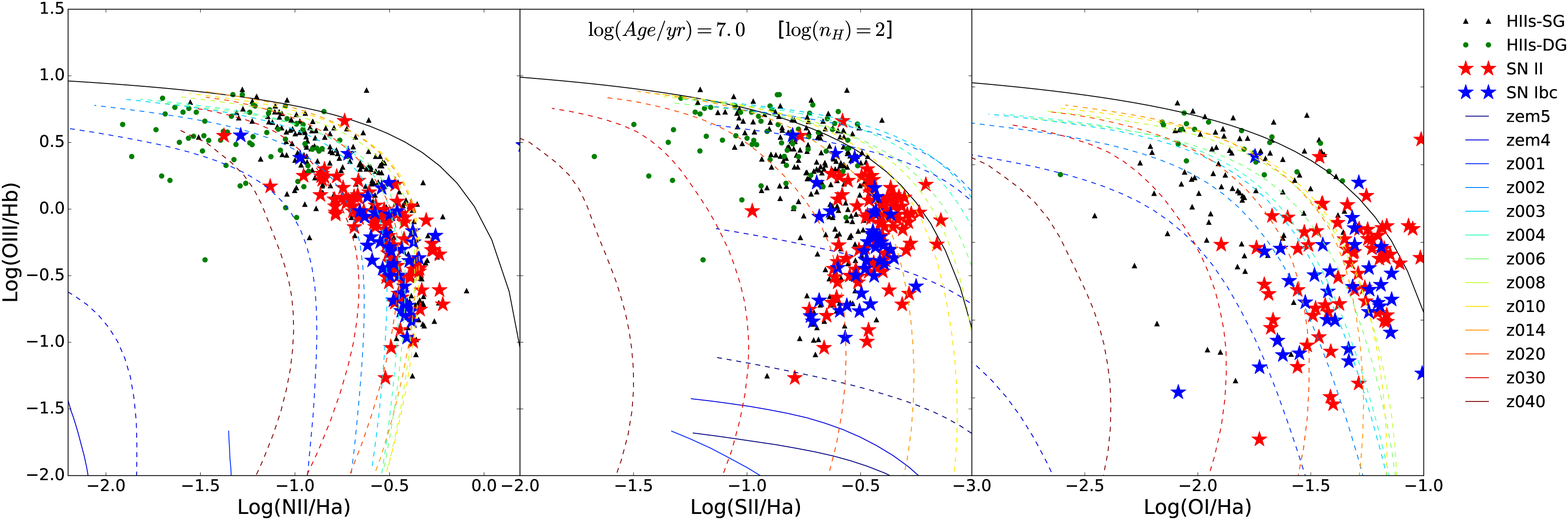} 
 \caption{The BPT diagrams at 2 different timescales separately from top panel to bottom panel, 3 Myrs, 7 Myrs and 10 Myrs. The tracks are in solid lines are single models and in dashed lines are binary models. Colour indicates metallicity, ranging from Z=0.0001 (blue) to 0.040 (red). The values of ionization parameter of the models is reduced following the track from upper left ($ log(U)= -1.5 $) to lower right ($ log(U)= -3.5 $).}
   \label{fig1}
\end{center}
\end{figure}

The BPT diagrams use strong optical emission line ratios, such as $ {\rm [OIII]\lambda 5007/H\beta} $, $ {\rm [NII]\lambda 6584/H\alpha} $, $ {\rm [SII]\lambda 6724/H\alpha} $, $ {\rm [OI]\lambda 6300/H\alpha} $, to probe the hardness of the ionizing radiation field in nebular emission regions (\cite[Baldwin et al.1981]{1981PASP...93....5B}). In Figure \ref{fig1} we show the BPT diagrams with $ {\rm \log(n_{H}) = 2 } $ to discuss the effect of binary evolution, metallicity and age on emission line ratios. 

We find the 13 different metallicity models produce their separated tracks in the BPT diagrams, extending from high log([OIII]/H$ \beta $) and low log([NII]/H$ \alpha $), log([SII]/H$ \alpha $) and log([OI]/H$ \alpha $) as the ionization parameter decreases. Both the highest and lowest metallicity models fail to match the distribution of observed HII regions. Models with metallicities ranging from $ {\rm Z = 0.001 } $ to 0.020 (roughly a twentieth Solar to Solar) pass through the observational data at ages of around 5 Myrs for both single and binary Models. In the single models this emission drops away quickly, while binary models continue to match the data through to ages of 10 Myrs.

\begin{figure}[h]
\begin{center}
\hspace*{-1.cm} \includegraphics[width=2.5in]{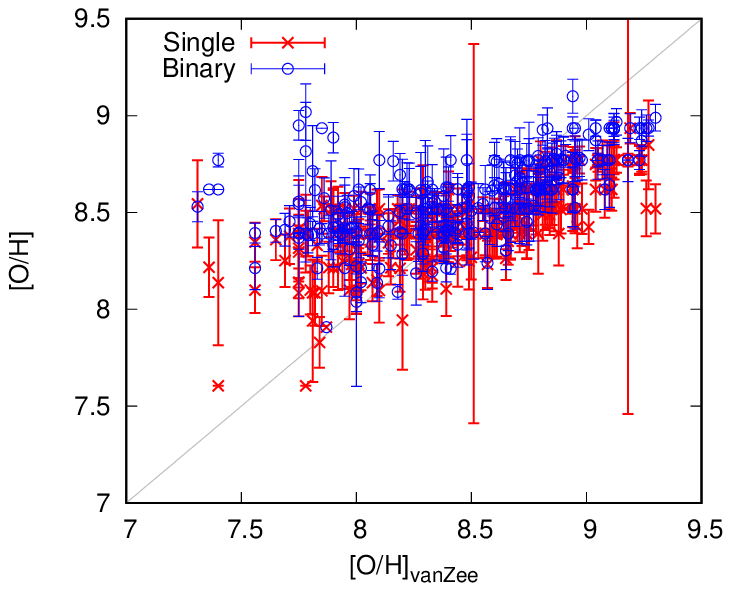} 
\includegraphics[width=2.5in]{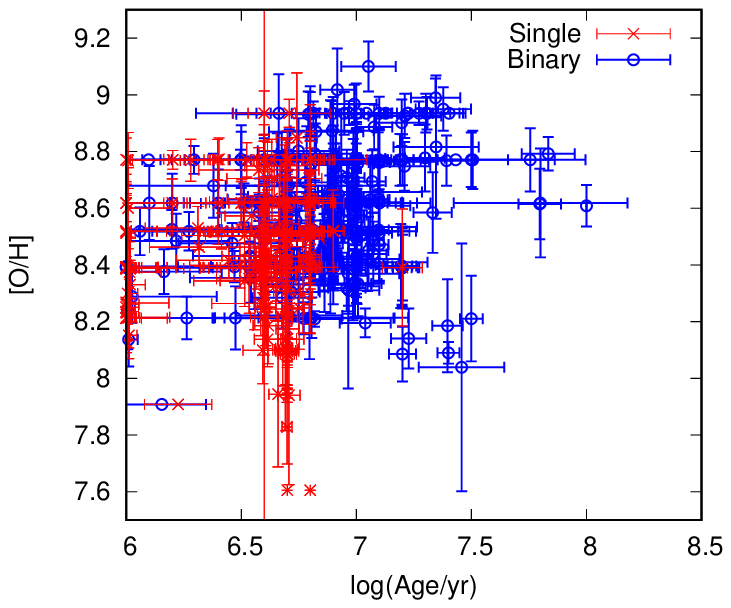} 
 \caption{Best-fitting models from single model, the red crosses with error bars, and binary model, the blue circles with error bars, for van Zee sample are presented here. The left panel shows [O/H] ($ {\rm 12 + \log{[O/H]}} $) estimation from single and binary models compared to the [O/H] from van Zee survey. The right panel presents age from these best-fitting models.}
   \label{fig2}
\end{center}
\end{figure}

The location of a model HII region in the BPT diagram depends on numerous factors. We find the best-fitting models to match these individual HII regions respectively by varying all of the input nebular parameters including age, hydrogen density, metallicity and ionization parameter. This allows us to derive the [O/H] and age of these observed HII regions. In the left panel of Figure \ref{fig2}, we compare our derived [O/H] metallicities with those determined by van Zee survey based primarily on strong line ratios. These are calibrated from photoionization models much like our own, but assuming a different relation between ionizing spectrum and stellar metallicity. We find that both our single and binary models suggest a narrower range of metallicities for these sources than the van Zee survey estimates. With both ends of the distribution moving back towards to sample mean. When the effects of binaries are included, our moderate metallicity models ($ 12 + \log{[O/H]} \sim 8.5 $) produce similar line ratios to those used by van Zee survey with $ 12 + \log{[O/H]} \sim 7.5 $.

Perhaps the most important property we derive for these HII regions is age. It provides an insight into the evolution history of massive stars and may also provide constraints on core-collapse supernova progenitors. In the right panel of Figure \ref{fig2} we present the estimation on age of these HII regions from the van Zee sample and a clear difference between single and binary model stands out. Firstly for these HII regions, binary model can extend to 30 Myrs, while most regions have ages around 10 Myrs. In comparison single-star models are younger within 10 Myrs and mostly around 4 Myrs. These results are consistent with the binary evolution pathway of massive stars which become cool supergiant stars due to mass loss, rather than hot helium stars as in single star evolutionary models.  

\begin{figure}
\begin{center}
\includegraphics[width=5in]{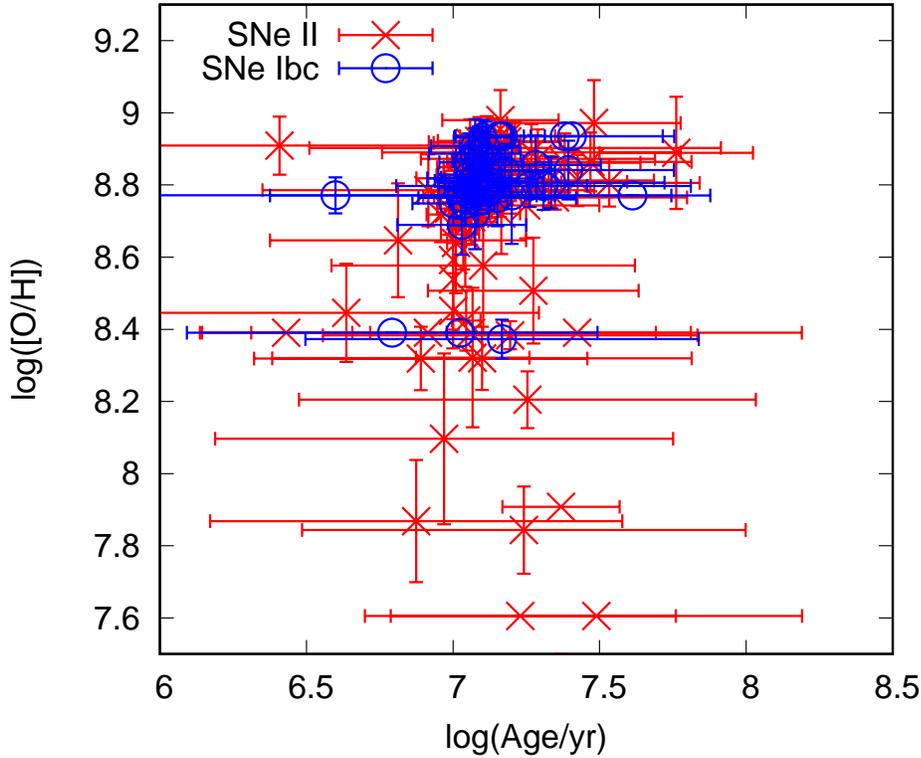} 
 \caption{The relation between [O/H] and age for SNe II and SNe Ibc from binary models. The red crosses with errorbars correspond to SNe II and blue circles with errorbars stands for SNe Ibc.  }
   \label{fig3}
\end{center}
\end{figure}

\section{Constraining the Progenitor Ages of Core-collapse 
Supernovae}
We can perform a similar analysis of the emission spectra found from stellar populations at the sites of core-collapse supernovae. Using the single-star models we find that the age of all core-collapse supernova progenitors would be less than 3 Myrs. In Figure \ref{fig3} we show the estimated ages from binary models for SNe II and SNe Ibc explosion sites. We find little difference in age for these two types of SNe. The important difference is that most of the ages for the supernova progenitors is around 10 Myrs or greater. This implies that the majority of progenitor stars for most supernovae have masses less than 20 $ M_{\odot} $ for both type II and type Ibc SNe. This is approximately the mass at which it is expected that core-collapse produces black-holes that may not lead to visible supernova (\cite[Eldridge \& Maund 2016]{2016MNRAS.461L.117E} and references therein). We note that there is also a strong metallicity preference for type Ib/c to arise from metallicities close to solar with most low metallicity events being of type II.

We caution these findings are still preliminary, but indicate that modelling the nebular emission from stellar populations including interacting binaries will provide a new method to understand core-collapse supernovae.

\end{document}